\documentclass [12pt]{article}
\usepackage[dvips]{color}
\usepackage{graphicx}
\usepackage{epsfig}
\usepackage{color}
\usepackage[normalem]{ulem}
\def\beq{\begin{equation}}
\def\eeq{\end{equation}}
\def\beqa{\begin{eqnarray}}
\def\eeqa{\end{eqnarray}}
\def\d{{\rm d}}

\begin{document}
\baselineskip0.6cm plus 1pt minus 1pt
\tolerance=1500
\vspace*{1cm}
\begin{center}
{\LARGE\bf
 Relativistic mechanical-thermodynamical formalism--description of  inelastic collisions}
\vskip0.8cm
{J. G\"u\'emez$^{a,}$\footnote{guemezj@unican.es},
M. Fiolhais$^{b,}$\footnote{tmanuel@teor.fis.uc.pt},
L. A. Fern\'andez$^{c,}$\footnote{lafernandez@unican.es}
}
\vskip0.3cm
{\it $^a$ Departament of Applied Physics, University of Cantabria, \\ {E-39005 Santander,  Spain} \\
\vskip0.3cm
{\it $^b$ Departament of Physics and CFisUC, University of Coimbra}, \\  P-3004-516 Coimbra, Portugal} \\
\vskip0.3cm
{\it $^c$ Departament of Mathematics, Statistics and Computation,  University of Cantabria, \\ E-39005 Santander,  Spain }
\end{center}

\vspace*{0.1cm}
\begin{abstract}
We present a relativistic formalism inspired on the Minkowski four-vectors that also includes conservation laws such as the first law of thermodynamics.
It remains close to the relativistic four-vector formalism developed for a single particle, but it is also related to the
classical treatment of problems that imperatively require both the Newton's second law and the energy conservation law.
We apply the developed formalism to inelastic collisions to better show how it works.
\end{abstract}

\section{Introduction}
\label{sec:introd}

It is not so uncommon to find in the literature papers whose motivation is to
link mechanics and thermodynamics \cite{leff93},
thus proposing ways to approach  textbook exercises from both sides  \cite{sherwood83}.
This approach obviously deals with Newton's law and with the conservation of energy as expressed by the first law of thermodynamics  \cite{guemez13a}.
In fact, some textbooks exercises, involving dissipation of mechanical energy or mechanical energy production,   are solely approached from the mechanical point of view    \cite{bauman92}.
Since some energy transfers or some energy variations are not described by Newtonian  equations,
such approaches give credit to those authors arguing that a satisfactory integration of classical mechanics and thermodynamics has not yet been achieved    \cite{arons99}.

Intuitively, one might think that the best way to integrate  mechanics and thermodynamics is by means of a theory in which both areas of physics are  imbedded and,
as defended in~\cite{margaritondo03}, Einstein's  theory of special relativity   is a good candidate. Indeed,
the theory of relativity must deal with all types of energies involved in a process and it associates linear momenta to these energies, a general feature imposed by the Lorentz transformation of a four-vector~\cite{rahaman14}.

Even if we were interested in processes occurring at low velocities, the special theory of relativity would offer the appropriate framework
to develop a formalism that integrates dynamics and thermodynamics on equal footing~\cite{guemez2014br}, therefore avoiding conceptual difficulties associated with that unification, as noted by some authors   \cite{lehrman73}. In the relativistic framework,
one has to regard the energy variations, the heat, and the work as components of four-vectors which obey a mechanical-thermodynamical fundamental equation    \cite{guemez10}.

Relativistic treatments of problems that require both a mechanical and a thermodynamical approach are not common in the literature \cite{landsberg81}
(a notable exception is the recent textbook by Chabay and  Sherwood   \cite{sherwood11}).
This paper is  a  contribution to incorporate relativity in  thermodynamics.
We start  with the special relativity formulation using four-vectors in the Minkowski space   \cite{gron80,essen02}
and incorporate, in that known formalism, conservation laws such as the first law of thermodynamics   \cite{guemez10,guemez13e}.
We illustrate how the formalism works in practice, using the example of a
deformable ball inelastically colliding  against a wall. This is the type of exercise involving extended macroscopic bodies that can be found in relativity textbooks   \cite{sears68}, with references to heat, temperature variations and internal energy.

The formulation of a relativistic thermodynamics has been a long and not yet successful process, but this paper is not the appropriate stage for  thoroughly  reviewing or criticizing those works. Nevertheless, we shall try to give a brief overview focusing on the disputes that most probably have been responsible for obstructing progress. In our opinion, the ``temperature transformation" has been one of those issues. In our presentation, based on the asynchronous formulation of relativity, it turns out that the temperature is the same in all inertial frames, i.e., the temperature is a Lorentz scalar.

This paper is organized as follows. In section \ref{sec:formalism} we present the essentials of the formalism, and explain how conservation laws can be embodied. Some
points in our review of the  Minkowski four-vector
formalism, such as the action of several forces acting on an extended body in a relativistic treatment, are presented with detail in the appendices. The synchronous and the asynchronous formulations of relativity are discussed in the context of the construction of a coherent  relativistic thermodynamic formalism.
In section \ref{sec:ineslascolliso}  we illustrate the formalism using the mentioned example of an inelastic collision of a ball against a wall.
This problem is analysed at constant temperature, in two different reference frames.
In section \ref{sec:adiinecoll} we examine a similar problem but now considering that the  mechanical energy variation in the collision process is reabsorbed by the body as internal energy (adiabatic process), so it experiences  a temperature (and an inertia)   change. The discussion of concrete examples, even academic ones, is usually absent from
the presentations of relativistic thermodynamics which are much focused on the formalism. By discussing those examples, we demonstrate that we have developed a formalism able to deal with
concrete situations, besides showing the formalism itself at work.
Section \ref{sec:conclus}  is devoted to the conclusions.

\section{The formalism}
\label{sec:formalism}

In a  already long series of papers we explored the complementary aspects of Newtonian mechanics and thermodynamics when one solves  textbook exercises involving extended systems, in particular rigid or deformable \cite{guemez13a} or articulated bodies    \cite{guemez13, guemez14,guemez14c}. This is carried on by exploring the equations
\beq
\Delta K_{\rm cm}=\int {\vec F}_{\rm ext} \cdot \d {\vec r}_{\rm cm}\,  \mbox{ \ \ \ \ \ (Newton's \ law)} \label{newtonptx}
\eeq
(usually expressed, in vector form, as $\Delta \vec P_{\rm cm}= \int  {\vec F}_{\rm ext} \, \d t$, where $ {\vec P}_{\rm cm}$ is the center of mass linear momentum) and
\beq
\Delta K_{\rm cm}+ \Delta U = W_{\rm ext} + Q \,  \mbox{ \ \ \ \ \ (First \ law \ of \ thermodynamics)}.\label{primllei}
\eeq
In these equations, $\Delta K_{\rm cm}$ is the variation of the center of mass kinetic energy; $\Delta U$ is the variation of the internal energy;  the work of the external forces,
$W_{\rm ext}$, and the heat, $Q$, in (\ref{primllei}), are the energy transfers crossing the system boundary.

In those papers, we particularly emphasized that the second member in (\ref{newtonptx}), the so-called pseudo-work, should not be confused with the real work: in general, for a system of particles, $\Delta K_{\rm cm}\not = W_{\rm ext}$. For each external force, ${\vec F}_{j}$, the corresponding work is given by $W_j= \int {\vec F}_j \cdot \d \vec r_j$, where the infinitesimal displacement $\d \vec r_j$ refers to the application point of that external force.  In (\ref{newtonptx}), the force in the integral is the resultant external force, ${\vec F}_{\rm ext}= \sum_j  \vec F_j$ (the resultant internal force vanishes, according to Newton's third law) and the infinitesimal displacement $\d \vec r_{\rm cm}$ refers to the center of mass of the system. In (\ref{primllei}), the total external work is $W_{\rm ext}= \sum_j W_j$.  In some cases, such as for a system consisting of a single particle, we may have $\Delta U =Q=0$ and the above mentioned distinct physical laws provide  the same information (this happens whenever the work is equal to the pseudo-work).

 In this article we aim at incorporating relativity in the formalism developed in that series of papers.

\subsection{Relativistic mechanics and thermodynamics}

In intermediate level courses on special relativity, one describes the dynamics of a single particle by an equation that is formally very similar to the Newton's second law, written as $\vec F={\d \vec p \over \d t}$, namely   \cite{freund08}
\beq
F^\mu={\d p^\mu \over  \d \tau} , \label{hlop}
\eeq
where $F^\mu$ is the four-vector force, $p^\mu$ is the four-vector momentum  and $\d \tau = \gamma^{-1} \d t$ is the infinitesimal  proper time interval,  $\gamma$ being the usual relativistic factor. In  Appendix A we explicitly write down the components of the four-vectors force and momentum.

An expression that resembles Newton's second law as expressed by equation (\ref{newtonptx}) can also be derived for describing the dynamics of a single relativistic particle. In differential form, such equation is written as
\beq
\d E^\mu=\delta W^\mu \, . \label{eqrelat}
\eeq

Equation (\ref{eqrelat}), whose similarity with the work-kinetic energy theorem expressed by  (\ref{newtonptx}) is rather obvious,
can be regarded as a
`momentum-energy / impulse-work equation' and it is indeed equivalent to the more familiar equation (\ref{hlop}). The four-vector differential energy on the left hand side of (\ref{eqrelat}) is  just $\d E^\mu=c\, \d p^\mu$, and it   is an exact differential. From (\ref{hlop}), we may write  $\d E^\mu = c\, \d p^\mu= c\, F^\mu \d \tau$ and, defining the infinitesimal impulse-work four-vector as $\delta W^\mu= c\, F^\mu \d \tau$ (the infinitesimal work is not an exact differential, therefore it is denoted by $\delta)$, one arrives at equation (\ref{eqrelat}). In Appendix A we give more details of the relativistic dynamics of a single particle, and then we generalize  equation (\ref{eqrelat}) to systems of particles.

Explicitly,  equation (\ref{eqrelat}) reads as   \cite{chrysos04}
\beq
\d E^\mu =
\left(\begin{array}{c} c \ m   \ \d \left[\gamma (v) v_x\right]\\ c \ m   \ \d \left[\gamma (v) v_y\right]\\ c \ m   \ \d \left[\gamma (v) v_z\right] \\ { m} \ c^2 \ \d \left[\gamma (v) \right] \end{array}\right) = \left(\begin{array}{c}  c\ F_x \ \d t\\ c \ F_y \ \d t\\ c \ F_z \ \d t \\  F_x\d x + F_y \d y + F_z \d z \end{array}\right)  = \delta W^\mu \, .
\label{explicit45}
\eeq
for a particle of mass $m$
and velocity $\vec v = (v_x,v_y,v_z)$, in reference frame S, acted upon by the force $\vec F = (F_x, F_y, F_z)$. The function $\gamma=\gamma(v)$  has its usual meaning:
\beq
\gamma(v)=[1-\beta(v)^2)]^{-1/2}\, , \ \ \  {\rm with}  \ \beta(v)={v \over c}\, .
\eeq

It is important to note that both $\d E^\mu$ and $\delta W^\mu$ in (\ref{explicit45}) are four-vectors, i.e., their components, under Lorentz transformations between inertial reference frames, transform like the components of the position-time contravariant four-vector $x^\mu$. It is also important to note that $\d t$ is a time interval measured in the reference frame S (by a set of two synchronized clocks) and should not be confused with the proper time $\d \tau$, entering in the definition (\ref{hlop}),  measured by a single clock that moves with the application point of the force, or, in other words, with the object that is a point-like particle. This distinction is crucial when one generalizes $\delta W^\mu$, as we shall do later on, to include the effect of several forces acting then on an extended body.

Going back to  (\ref{explicit45}) and to the point-like particle, the set of the first three equations  --- the space-like components --- can be regarded as the relativistic counterpart of Newton's second law in vector form (i.e.
corresponding to the non-relativistic equation of motion $m \, \d {\vec v}={\vec F} \, \d t$) \cite[p.~277]{freund08}; and the equation for the time-like fourth component can be regarded as the equation corresponding to the differential form of the non-relativistic equation~(\ref{newtonptx}) for the single particle case
(i.e. corresponding to $\d K = {\vec F} \cdot \d {\vec r}$). It is worth noticing that  this relativistic time-like equation can be obtained  from the top three space-like equations
by using  (see Appendix B for the proof of this identity)
\beq
{\d [\gamma (v)  c^2] } =  v_x \, \d [\gamma (v) v_x]  + v_y \, \d [\gamma (v) v_y]  + v_z \,  \d [\gamma (v) v_z] \, . \label{er4tsdi}
\eeq
This relation, that establishes the equivalence  between the information provided by the
space-like components and by the
time-like component of equation (\ref{explicit45}), is the relativistic counterpart of the expression  $  \d [{1\over 2}v^2]  = v_x \d v_x  +  v_y \d v_y +  v_z\d v_z$,   that allows the derivation, in classical mechanics,  of the work -- kinetic energy theorem (\ref{newtonptx}) (of course,  that equation also applies to a single particle) starting from Newton's law in vector form.
Hence, in a sense, there is some redundancy in the information provided by the set of four components in the four-vector equation~(\ref{explicit45}).
However, this redundant information, inherent to equation (\ref{explicit45}), is not present, in general, when we generalize the formalism to systems of particles, such as  extended deformable bodies.

In the spirit of equation (\ref{primllei}), the generalization of expression (\ref{eqrelat})
for an extended  body acted upon by various forces and undergoing a process in which non-mechanical energies are present
is  \cite[p.~221]{sherwood11}
\beq
\d E^\mu=   \delta W^\mu + \delta Q^\mu \label{eqrelat2}
\eeq
with $\delta W^\mu = \sum_j \delta W_j^\mu $ being the impulse (space-like part) and the  work (time-like part) of the external forces acting on the system, and $\delta Q^\mu$ is a four-vector associated with the energy exchanged as heat   \cite{kampen68} to be discussed later on.

When an extended body (or a system of particles) is considered, there is an
internal energy associated to the system.  The internal energy in the rest frame of the body, $U$, is related to the inertia of the body, $M$, through   \cite{hecht11} $U=M \, c^2$
or $M=Uc^{-2}$ which, being the same expression, better expresses the idea that the inertia of the system comes from its internal energy.
Though in thermodynamics it is not required to make any microscopic hypothesis about the constitution of a system, it is tempting to do so and to relate the internal energy of the body  with the  kinetic energies of its constituents, in the reference frame where the system is at rest, as well as with the potential energies
associated with all the interactions {\em inside} the body. The clustering of the particles (whatever they are) forming the system leads to an energy decrease  with respect to the configuration where all the constituents of the body are at rest and far away from each other --- the binding energy ought to be negative  \cite{plakhotnik06}. On the other hand, any temperature increase always leads to an internal energy increase \cite{marx91}.
If the system, in its rest frame, is assumed to be made out of elementary particles (electrons, quarks, the Higgs boson, whatever they are), the internal energy can always be expressed by  \beq
U(T)=\sum_i m_i \, c^2 - \tilde{U}+\int _{T_0}^T C_P(T) \, \d T= U+\int _{T_0}^T C_P(T) \, \d T
\label{internfgt}
\eeq
where $m_i$ is the inertia of each constituent particle, $-\tilde U$ is the binding energy, $U$ is the internal energy at some reference temperature, $T_0$, and  $C_P$ is the heat capacity of the body.

In the reference frame where the system is at rest, $U$ is the fourth component of the energy four-vector, and it is the only non-vanishing component of that four-vector. In another inertial frame, there are space-like components different from zero, as is happens for the single particle case. Moreover, if the body of inertia $M$ is moving with velocity $v$ with respect to some inertial frame, the kinetic energy of the body is $K=\left[\gamma (v) -1\right] U$ or, equivalently, $E=K+Mc^2=K+U$, where $E$ is the energy of the body [by writing the energy in this form, it is directly relatable with the left-hand side of (\ref{primllei})].
Of course, the inertia of a composite body is not an absolute  constant because it may change. In particular it changes when the temperature of the body varies, when its composition gets modified, etc., as expressed by equation (\ref{internfgt}). Hence, the inertia, $M$, depends on the temperature but, since it is a relativistic invariant, all observers, in any inertial frame, must assign the same temperature to the body   \cite{landsberg96}.  This statement is inherent to our treatment, i.e.  it is not an {\em ab initio} assumption. Rather, it is a consequence of the invariant norm of a Minkowski four vector  --- the inertia ---, which is directly related to the energy-momentum of the system. In subsection \ref{ssec:asynchrform} we shall discuss the Lorentz scalar nature of the temperature, then in connection with a photon gas system.

If there are several external forces acting on the relativistic body, we introduce the following four-vector
\beq
\delta W^\mu= \sum_j \delta W_j^\mu= \sum_j \left(\begin{array}{c}  cF_{j,x} \d t \\ cF_{j,y} \d t \\ c F_{j,z} \d t \\  F_{j,x}\d x_j + F_{j,y} \d y_j + F_{j,z} \d z_j \end{array}\right) \label{traghds}
\eeq
where $\vec F_j$ represents each external force and the differentials are the components of the four-vector infinitesimal displacement
$\d x_j^\mu = (\d x_j,\d y_j,\d z_j,c \d t)$ (note that the time interval in S is the same for all forces  \cite[p.~251]{sherwood11}). In the Appendix A we give more details about the generalization that allows us to write equation (\ref{traghds}).  For the same time interval $\d t$ in the space-like components of (\ref{traghds}), each term $\delta W_j^\mu$ is a four-vector (a proof is given in Appendix C) and therefore $\delta W^\mu$ is a four-vector indeed: in any other inertial frame, the  components of $\delta {W '}^\mu$ are obtained after the application of the Lorentz transformation matrix to (\ref{traghds}). The subtle point is the requirement of $\d t$ to be the same in S for all forces. This means that the proper time $\d \tau_j$ relative to the force $j$, is generally different from the other proper times for the other forces.

Let us denote by S$'$ an inertial  reference frame that moves from left to right with velocity $V$ along the $x$ axis, i.e. a reference frame in standard configuration \cite{gron87}.
The Lorentz matrix transformation
readily allows us to convert any four vector and, therefore, any four-vector equation, from one inertial reference frame to another one   \cite{jefimenko97}.
For the standard configuration,   the transformation matrix   is given by
\beq
{ \Lambda}_\nu^\mu  (V) =\left(\begin{array}{cccc} \gamma (V)   & 0 & 0 & -\beta (V) \gamma (V)  \\0 & 1 & 0 & 0 \\0 & 0 & 1 & 0 \\-\beta (V)  \gamma (V) & 0 & 0 & \gamma (V) \end{array}\right)\, .
\label{eq-1}
\eeq
When it is applied to equation (\ref{eqrelat2}), this leads to   \cite{gamba67}
\beq
{ \Lambda}_\nu^\mu  (V) \left[ \d E^\nu=\delta W^\nu + \delta Q^\nu \right] \ \ \ \Leftrightarrow   \ \ \  \d E'^\mu=\delta W'^\mu + \delta Q'^\mu \, . \label{eqrelat3}
\eeq
Similar transformations can be applied to any other four-vector or four vector equation.

Going back to equation  (\ref{eqrelat2}), $\delta Q^\mu$ stands for the four-vector heat transferred  from a reservoir  to the system or from the system  to the reservoir.
In a reference frame where the heat reservoir is at rest,  the  only non-vanishing component is the fourth one, i.e. the three-momentum associated with the heat should be zero in that particular frame    \cite{guemez10}.
If we take the example of heat transfer as a process being associated with the emission or absorption of photons,
the  corresponding overall linear momentum is zero   \cite{field14}  (i.e., for each photon which is emitted in one direction    in average there is another photon, of the same frequency, emitted in the opposite direction. This is what we take into account when we write down $\delta Q^\mu$ as
\beq
\delta Q^\mu = \left(\begin{array}{c}  0 \\ 0 \\ 0 \\  \delta Q \end{array}\right)\, . \label{4calor}
\eeq
However, the vanishing of the space-like components of the heat four-vector is not general. In a reference frame where the heat reservoir is not at rest, there should
be a space-like component for $\delta Q^\mu$ and this is the impulse associated with the heat transfer. The relativistic Doppler effect and the aberration effect   \cite[Chap.~31]{freund08}
provide the explanation for this result. This `non-mechanical' impulse plays a role similar to the normal impulse of the resultant of the external forces, since both contribute to change the three-momentum of the system. Therefore, equation  (\ref{eqrelat2}) presents, simultaneously, the conservation laws for the energy and for the linear momentum: the energy of a system varies as a consequence of work or heat crossing the system boundary  and the linear momentum of the system varies because of the impulse (of mechanical and non-mechanical origin) crossing the same boundary.

As discussed in the Appendix A, the heat itself, as the counterpart of the heat in thermodynamics,  should be regarded as the norm
$\left|\left|\delta Q^\mu\right|\right|$ of the four-vector   (\ref{4calor}). This norm  is $\delta Q$ for the four-vector $\delta Q^\mu$ and it is a relativistic invariant   \cite{yuen70}.

In the next section we examine, through  an  example, the usefulness and the predictions that may be obtained from  equation~(\ref{eqrelat2}).
 But, before that,  in the next subsections we shall discuss several aspects related to the construction of a relativistic thermodynamical theory.
\vskip0.5cm

\subsection{Brief review of relativistic thermodynamics --- synchronous and asynchronous formulations}
\label{ssec:brrevwrelther}

What usually is mentioned as {\em relativistic thermodynamics} is not the proposal of relativistic equations suitable to be applied to thermodynamics problems \cite{landsberg81}, but rather the search
for the relativistic transformations of thermodynamical magnitudes. This mainly applies to the temperature, whose transformation rule has been a matter of dispute \cite{landsberg96}, though there is no evidence of any experimental methodology proposed to distinguish between the various options \cite{landsberg80}.
Once the first principle of thermodynamics is written using four-vectors, such as
 $\Delta E^\mu = W^\mu + Q^\mu$ \cite{kampen68},
to the extent of our knowledge, concrete problems have only been addressed in the framework of that equation in a previous paper by one of the authors
\cite{guemez10}.

Briefly, one may say that the formulation of a relativistic thermodynamics was approached from two different sides: the so called
`synchronous formulation' \cite{cavalleri78}
and the so-called
`asynchronous formulation' \cite{cavalleri69}.
Let us imagine an experiment with some simultaneous events in S, such as a set of forces acting simultaneously, at different positions, upon a system. Let us then imagine the experiment with the very same simultaneity character in  S$'$, i.e. a second but an exactly similar experiment repeated in S$'$. When we want to relate both experiments we are on the grounds of the `synchronous' formulation of relativity.
The term `asynchronous' applies to when an experiment, whose different parts are simultaneous in S, is then also observed (now  necessarily non-simultaneously) in  S$'$ --- in this case  one has just a single experiment \cite{gamba67}.
The experiment is described in S by certain coordinates and magnitudes, but it is also observed in S$'$ where it is described with different coordinates and magnitudes
 \cite{goodinson85}.
In many discussions and comments, the authors seem to not completely realize that, when they are defending their view points, they are talking about distinct formulations of relativity.

Let's make even more clear the distinction between the synchronous and the asynchronous formulations. An observer in S performs an experiment in which two events  take place simultaneously: they are described in S by the four-vectors ${x}_1^\mu = \left(x_1, 0, 0, t \right)$ and ${x}_2^\mu = \left(x_2, 0, 0, t \right)$. A similar experiment is led in S$'$, in standard configuration with velocity $V$ with respect to S, imposing that the events should also take place simultaneously
\cite{gron81}.
The two four vectors describing the corresponding events are ${x'_1}^\mu = \left(x'_1, 0, 0, t' \right)$ and ${x'}_2^\mu = \left(x'_2, 0, 0, t' \right)$.
The experiments in S and in S$'$ are {\em distinct} and therefore ${x}_1^\mu$ and ${x'}_1^\mu$ are not related by a Lorentz transformation, i.e. ${x'}_1^\mu \ne {\Lambda}_\nu^\mu (V) {x}_1^\nu$ and,
of course, the same applies to ${x}_2^\mu$ and ${x'}_2^\mu$.

As far as the asynchronous formulation is concerned
\cite{gron73},
an observer carries on an experiment, such that two events take place simultaneously in S. For instance he observes two forces, $F_1$ y $F_2$,
simultaneously applied in the space-time intervals $\d x_1^\mu = \left(\d x_1, 0, 0, c\d t\right)$ for $F_1$ and
$\d x_2^\mu = \left(\d x_2, 0, 0, c\d t\right)$, for $F_2$. In S, the associated impulse-work four-vectors are
$\delta W_1^\mu = \left(c F_1 \d t , 0, 0, F_1 \d x_1\right)$ and $\delta W_2^\mu = \left(c F_2 \d t , 0, 0, F_2 \d x_2\right)$, respectively.
Now, an observer in S$'$ in standard configuration does not conduct a similar experiment but rather expresses, in his own space-time coordinates, the events and the space-time intervals associated with the experiment. In S$'$ the corresponding space-time intervals are the following four-vectors:
$\d {x'}_1^\mu = \left(\d {x'}_1, 0, 0, c\d {t'}_1\right)$ and $\d {x'}_2^\mu = \left(\d {x'}_2, 0, 0, c\d {t'}_2\right)$.
Clearly, events and processes that are simultaneous in S, will not be simultaneous in S$'$  (relativistic non-simultaneity effect), justifying the denomination `asynchronous'.
 One postulates that the four-vectors in S$'$ are related to the S ones by the Lorentz transformation, e.g. $\d {x'}_1^\mu  =  {\Lambda}_\nu^\mu (V) \d {x}_1^\mu$, (the same for $\d {x'}_2^\mu$  and  $\d {x}_2^\mu$).
In general, in the asynchronous formulation, any four-vector in S, is expressed in S$'$ by its corresponding transformed four-vector, ${A'}^\mu$. This means that the same process described in S by certain coordinates and magnitudes is now described in S$'$ by the coordinates and magnitudes of this reference frame. As for any four-vector, one has:
\beq
{A'}^\mu  =  {\Lambda}_\nu^\mu (V) A^\nu\, ; \ \ \ {\rm and} \ \ \ {A}^\mu  =  {\Lambda}_\nu^\mu (-V) {A'}^\nu \, .
\eeq

\subsection{The asynchronous formulation}
\label{ssec:asynchrform}

In our perspective, the asynchronous formulation of relativity provides an appropriate ground  to develop a relativistic thermodynamic formalism and it is the one adopted in this article. In this framework we are able to describe not only pure thermodynamical processes but also those processes involving dissipative forces whose description requires both mechanics and thermodynamics. The methodological process is clear: first, one has to construct the Minkowski four-vectors associated with the process which is described by equations between these four-vectors; then we may use the Lorentz transformations to relate the observations of the {\em same} process in one and in any other inertial reference frame, therefore enforcing {\em ab initio} the fulfilment of the  first postulate of the Einstein's relativity theory.

For the asynchronous formulation, a process is described in a given reference frame, say the inertial frame S, by the four-vector equation
 $E_{\rm f}^\mu - E_{\rm i}^\mu = \sum_j W_j^\mu + Q^\mu$ \cite{kampen69}
such that (i) all forces are simultaneously applied during the same time interval, $\Delta t$, though they might be applied at different points of the system and with different displacements;
(ii) the heat reservoir is at rest in S and  there is no net linear momentum associated with the heat.
The corresponding equation in S$'$, i.e. the one written by an observer in S$'$ for the same process is obtained in a straightforward way just by applying the Lorentz transformation to the above four-vector equation~\cite{rothenstein95}:
\beq
{\Lambda}_\nu^\mu (V)  \left[E_{\rm f}^\nu - E_{\rm i}^\nu = \sum_j W_j^\nu + Q^\nu \right] \rightarrow {E'}_{\rm  f}^\mu - {E'}_{\rm i}^\mu = \sum_j {W'}_j^\mu + {Q'}^\mu\, ,
\eeq
which is the integral form of equation (\ref{eqrelat3}) when several simultaneous forces on S are applied to the system.
This asynchronous formulation guarantees the fulfillment of the principle of relativity, since the equations are covariant under Lorentz transformations, even though forces simultaneously applied from the point of view of S are not simultaneous in S$'$, and a set of thermal photons with zero linear momentum in S must have a linear momentum different from zero in S$'$. It also guarantees the fulfilment of the Einstein equation, in the sense that the  internal energy of the system totally contributes to its inertia. Finally, the asynchronous formulation
guarantees that, in the limit $c \rightarrow \infty$, the non-relativistic equations --- Newton's second law and the first law of thermodynamics --- are recovered, a necessary condition for the consistency of any relativistic theory. In particular, in the non-relativistic limit, the forces are simultaneously applied in all reference frames.

In the asynchronous formulation of the relativity, the relativistic transformation of the quantities that are components of a four-vector
is prescribed by the Lorentz transformation. However, if a physical magnitude is not directly related to those components, such as the temperature, the transformation properties can be indirectly obtained.
For the sake of illustration, let us consider  an ensemble of thermal photons contained in a cavity of volume ${\cal V}$ at rest (in reference frame S) in thermodynamical equilibrium at temperature $T$~\cite{yuen70}. The global linear momentum is zero in reference frame S, meaning that, in average, for a photon  moving in one direction $(\theta, \phi)$ there should exist another one of the same frequency moving in the opposite direction. The internal energy of the system is $U(T)=a{\cal V} T^4$, where $a$ is the so-called radiation constant. The energy-momentum four-vector of each photon is
$
e^\mu_n =  \left\{ c(h\nu_n/c) \cos \theta_n \cos \phi_n, c(h\nu_n/c) \sin \theta_n \cos \phi_n, c(h\nu_n/c)   \sin \phi_n, h \nu_n\right\}$,
and the sum of all these four-vectors, for the same instant in S, is the following four-vector
\beq
E^\mu = \sum_n e^\mu_n= \left(\begin{array}{c}  0 \\ 0 \\ 0 \\  \sum_n h \nu_n \end{array}\right) =
\left(\begin{array}{c}  0 \\ 0 \\ 0 \\  ANT \end{array}\right)
\, , \label{fotones}
\eeq
where $A$ is a universal constant and $N\sim {\cal V}\, T^3$ is the number of photons \cite{shanks56}.

For an observer in S$'$, the same photons have different frequencies, due to the Doppler effect, and they move in different relative directions, with respect to S, due to the aberration effect. For the observer in S$'$ the frequency distribution in not Planckian any more~\cite{landsberg04}.
The Lorentz transformation applies both to each photon,
${e'}_n^\mu = {\Lambda}_\nu^\mu (V) {e}_n^\nu$, and then one sums up, or it applies to the global four-vector. The result is the same:
 \beq
{E'}^\mu = {\Lambda}_\nu^\mu (V) {E}^\nu = {\Lambda}_\nu^\mu (V) \sum_n e_n^\nu =  \left( \begin{array}{c} - c \gamma V \left(c^{-2}\sum_n h \nu_n\right) V \\ 0 \\ 0 \\
\gamma  \left(\sum_n h \nu_n\right) \end{array}\right) =
\left(\begin{array}{c} - \beta \gamma ANT  \\ 0 \\0 \\ \gamma ANT \end{array} \right) \, ,
\eeq
where $\gamma=\gamma(V)$ and $\beta=\beta(V)$. If the observer in S$'$ computes the norm of the four-vector ${E'}^\mu $ he is bound to conclude that
$
\left|\left|{E'}^\mu \right|\right| =  ANT\, .
$
Recognizing that this is the internal energy of the system of $N$ photons in S$'$, the same as observed in S, he is also bound to assign the temperature $T$ to the system, exactly as in S.
In this formulation, the temperature is a Lorentz scalar but it is not the norm of any four-vector.
However, one should note that different formulations of relativistic thermodynamics may lead to a different conclusion on the temperature transformation \cite{yuen70}, an issue not yet  settled from the experimental point of view. Therefore, one may not strongly argue that the above result is unquestionable (though that is not our focus in this article).
But it is certainly an outcome in the framework of the adopted perspective and approximations.

\section{Inelastic collision}
\label{sec:ineslascolliso}

\begin{figure}[hbt]
\begin{center}
\hspace*{0.0cm}
\includegraphics[width=9.0cm]{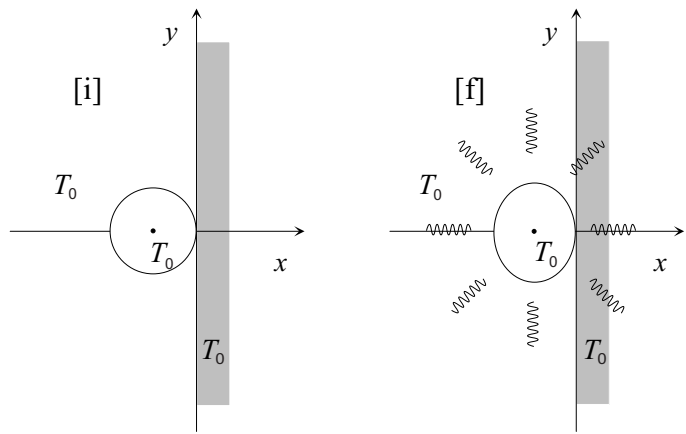}
\end{center}
\vspace*{-0.6cm}
\caption[]{\label{fig:1} \small A ball colliding with a wall at rest in reference frame S in an  isothermal process.}
\end{figure}

To  illustrate the formalism, let's take a ball, far from gravitational fields, moving with velocity $\vec v$ in the positive direction of  the $x$ axis, when it collides with a wall (of infinite mass) placed along the $y$ axis, as shown in Figure \ref{fig:1}   \cite{arons89}.
The wall is at rest in the reference frame S (the reference frame represented in the figure).

The ball has a positive electric charge $q$ at its center, and moves in a static electric field, $\vec E= E \vec e_x$, in reference frame S, that we assume to be also uniform, for the sake of simplicity. Using the formalism presented in the previous section and in the Appendix A we describe, in   S, the inelastic collision of the ball  since the instant  it touches the wall, until it stops.

During the collision process, there are two forces applied to the ball, namely, the electric force along the $x$ direction, of magnitude $F_{\rm e}=qE$, and the contact force in the opposite direction, $\vec N = -N   \vec e_x$ which is time dependent and responsible for the braking process. In the initial state, $N=0$ and in the final state $N=F_{\rm e}$.

We have to make some assumption on the thermodynamical character of the collision, and a simple choice is to consider it as an isothermal process as represented in Fig.~\ref{fig:1}: conceptually, we may think that the process evolves slowly enough for the dissipated mechanical energy might be emitted as heat and absorbed by the heat reservoir. Hence the ball keeps its temperature, $T_0$, which is also the temperature of the wall and of the involving surrounding that act as a static heat reservoir, in S, hence everything is at temperature $T_0$. Under such conditions, the inertia of the system, $M$, is assumed to be constant during the process.  Of course, due to the deformation effect, $\tilde U$ in equation
 (\ref{internfgt}) is not the same for the spherical and deformed ball, but they should not be very different specially if the deformation is not sizeable. For simplicity,  we assume $M$ to be the same, i.e. the deformation does not introduce an inertia variation,  and this is clearly an approximation in this study.
 The velocity of the ball, say, of its center of inertia   \cite[Sec.~6.7]{ferraro07}
is denoted by $v$, with   initial value $v_0$ and zero final value. Of course, the two forces are acting simultaneously and,
in the present case, equation (\ref{eqrelat2}) explicitly reads as
\beq
\left( \begin{array}{c} c \ M   \ \d (\gamma v ) \\ 0 \\ 0  \\ M \ c^2 \ \d (\gamma)  \end{array} \right) = \left( \begin{array}{c}  c F_{\rm e}  \, \d t\\ 0 \\ 0 \\ F_{\rm e}\, \d x   \end{array}\right)+ \left( \begin{array}{c}  - c N  \, \d t\\ 0 \\ 0 \\ - N\,  \d x_N \end{array}\right) + \left(\begin{array}{c}  0 \\ 0 \\ 0 \\  \delta Q \end{array}\right),
\label{explighyt}
\eeq
where $\d x$ denotes the displacement of the application point of the electric force, and $\d x_N$ the displacement of the application point of the contact force. Clearly, this stopping force does not displace its application point, and therefore $\d x_N =0$ (in other words, this force does not perform work)
and so we can write the following equations:
\beq
\left\{ \begin{array}{l} c \, M \ \d (\gamma v)=  c\, (F_{\rm e}-N) \d t \\ M\, c^2 \ \d \gamma = F_{\rm e}\, \d x+ \delta Q \, . \end{array} \right.  \label{duascois}
\eeq
Even before carrying on integrations, and making use of  equation (\ref{er4tsdi}), which in the present kinematical situation reads as
 $\d [\gamma c^2]/\d [\gamma v]=v$, the first of these equations can be written  in the following form:
\beq
M\, c \ \d (\gamma v)= c\, (F_{\rm e}-N) \d t  \Leftrightarrow  M c^2 \d \gamma = (F_{\rm e}- N) v \d t \Leftrightarrow M c^2 \d \gamma = (F_{\rm e}- N) \d x \label{ghjjky}
\eeq
where we have used $\d x= v \, \d t$ for the infinitesimal displacement of the object. Note that $v$ is a time dependent function.

The integration of the above equations requires a model for the stopping force. Let us assume that this is
a constant force of magnitude $\bar N$. Denoting by  $t_0$ the collision time (proper time for the application point of $\vec N$ since it does not move) we can view this average force as  ${\bar N}= t_0^{-1}\int_0^{t_0} N(t) \d t$. On the other hand, the initial and final energy four-vectors are  $E^\mu_{\rm i}$ and $E^\mu_{\rm f}$,  explicitly given by
\beq
E^\mu_{\rm i} = \left( \begin{array}{c}   c \, \gamma (v_0)\, M\,   v_0  \\ 0 \\ 0   \\ \gamma (v_0)M\, c^2    \end{array} \right) \ \ \ \
E^\mu_{\rm f} =   \left( \begin{array}{c}  0 \\ 0 \\ 0   \\ M\, c^2    \end{array} \right) \, .
\eeq
As mentioned above, we are assuming an isothermal process, i.e. the inertia of the system is the same before and after the collision and the heat is completely transferred  to the heat reservoir.
Now we are ready to integrate (\ref{duascois}), yielding
\beq
\left\{ \begin{array}{l}  - c \ \gamma (v_0)\, M\,   v_0 = c\, (F_{\rm e}-\bar N)  t_0 \\ M\, c^2 \ [1- \gamma(v_0)] = F_{\rm e}\Delta x + Q \end{array} \right. \label{duascoisas}
\eeq
The integration of (\ref{ghjjky}) is straightforward leading to
\beq
M\, c^2 \ [1- \gamma(v_0)]=   (F_{\rm e} - \bar N) \Delta x   \label{mmmjk}
\eeq
which, combined with the last equation in (\ref{duascoisas}), allows us to conclude that
\beq
Q =-\bar N \Delta x\, . \label{ghsy}
\eeq
The heat released in the process is equal to the pseudo-work performed by the stopping force.
On the other hand, from   equation (\ref{mmmjk})   heat can also be regarded as the variation of   kinetic energy of the ball and the work done by the electric force.
The result (\ref{ghsy})
is identical to the one obtained in the corresponding  non-relativistic collision~ \cite{guemez13a}. In particular, the variation of the entropy of the universe is positive also here,  $\Delta S_{\rm U}= - Q/T >0$,  so this process is irreversible. Moreover, in the limit $c\rightarrow \infty$, equations (\ref{duascoisas})  reduce to the corresponding  equations for that classical inelastic collision. The same happens, of course, with (\ref{mmmjk}) that reduces, in the same limit, to the center of mass equation (\ref{newtonptx}), i.e. to $-{1 \over 2} M v_0^2 = (F_{\rm e}-\bar N) \Delta x$.

\subsection{Principle of relativity}
\label{ssec:prinrelativ}

It is useful to look at the same process from the reference frame S$'$  in standard configuration with velocity $V$.
The Lorentz matrix transformation
applied to equation (\ref{eqrelat2}) [see equation (\ref{eqrelat3})]
 leads explicitly to
\beqa
& & \hspace{-2cm} \left( \begin{array}{c}  \gamma (V)\, c \ M\,  \d (\gamma v) - \gamma (V) \beta (V) M \, c^2 \d \gamma \\ 0 \\ 0 \\
 - \gamma (V)\,\beta (V) \, c \  M\, \d (\gamma v) + \gamma (V)  M \, c^2 \,  \d \gamma  \end{array} \right)  =   \nonumber \\
 & = &
 \left( \begin{array}{c}  \gamma (V) \, c \, (F_{\rm e}-N) \, \d t - \gamma (V) \beta(V) F_{\rm e} \d x \\ 0 \\ 0 \\ -\gamma (V) \beta (V) \, c \, (F_{\rm e}-N) \, \d t + \gamma (V)
 \, F_{\rm e} \, \d x
 \end{array} \right) +
 \left( \begin{array}{c} - \gamma (V) \beta (V) \, \delta Q \\ 0 \\0 \\ \gamma (V) \, \delta Q
 \end{array} \right) \, , \label{sdfrt}
 \eeqa
 where   $\gamma$ without any argument denotes $\gamma(v)$.
The nice feature with this global Lorentz transformation applied to equation (\ref{eqrelat2}) is that we do not have to bother about any transformation of the variables (such as the collision time),    or even about the transformation of the velocities: everything is properly taken care by the Lorentz transformation itself. However, it is interesting to explicitly check this point. Firstly, one recognizes that the left-hand side of  equation (\ref{sdfrt}) can be written  in the following form [see left-hand side of equation (\ref{explighyt})]:
\beq
\left( \begin{array}{c}  c\,  M  \ \d [\gamma(v') v'] \\ 0 \\ 0 \\
           M \, c^2 \d [\gamma(v')]  \end{array} \right)
\eeq
where
\beq
v' ={ v-V \over 1 - {v \, V / c^2}}
\eeq
[note that $\gamma (v') v' = \gamma (v) \gamma (V) (v-V)$ and $\gamma (v')   = \gamma (v) \gamma (V) (1 -vV/c^2)$]
is the well-known velocity transformation. Regarding the time interval during which the forces are applied, it is the same in S, and equal to $t_0$, but in S$'$ one has   $t'_{0,F_{\rm e}}   = \gamma (V) [t_0 - (V/c^2)  \Delta x]$ for $F_{\rm e}$, and $t'_{0,\bar N} = \gamma (V)  t_0$ for ${\bar N}$, i.e. forces simultaneously applied in S are not simultaneous  in S$'$   \cite{gron73}
But, as mentioned above,  the application of the Lorentz transformation as in (\ref{eqrelat3}) implies that everything is consistently taken into account.
Since the electric field in S is along the $x$ direction, the electric field in S$'$ is an identical vector and therefore the electric forces  in S and in S$'$ are the same, $F_{\rm e}$ \cite[p.~282]{freund08}, and there is no magnetic field in S$'$ either.

It is worth stressing the effects of the Lorentz transformation on the four-vectors on the right-hand side of  equation (\ref{eqrelat2}), i.e. on the momentum-energy transfers.
On the one hand, in the four-vector $\delta W^\mu$,  both the space- and the time-like components get modified. In particular there is now a work  $\delta W'_{\rm N}= \gamma (V) {\bar N} V \d t$  assigned to the contact force in S$'$ (the work of $\bar N$ is zero in S). On the other hand,    the four-vector $\delta Q^\mu$ acquires a space-like component along the $x$-axis (which is zero in S) that  is given
by $\d p'_Q= -\gamma (V) {\delta Q \over c^2} V$, where the appearance of the inertia associated with the heat   \cite{gabovich07}
${\delta Q \over c^2}$, is rather evident moreover, the time like
component of $\delta Q^\mu$ gets modified as well, with respect to S. In particular, for the heat, this means that, in S$'$, it does not flow isotropically, as it is the case in reference frame S.
A physical interpretation can be easily provided if we relate the heat transfer with the emission of thermal photons (i.e., an ensemble of photons with zero total momentum in S). Indeed, as already mentioned in section 2, the above result can be obtained by applying  the Doppler effect   \cite{saknidy85}
and the aberration effect transformations in S$'$   \cite[Sec.~7]{guemez10}. In particular, $\delta Q/c^2$ turns out to be  the inertia associated with the ensemble of thermal photons \cite{kolbens95}
and it is a Lorentz invariant.
The matrix equation (\ref{sdfrt}) reduces to the set of equations
\beq
\left\{ \begin{array}{l} c \ M \,  \d (\gamma v) - \beta (V) \, M c^2 \, \d \gamma = c (F_{\rm e}-N) \d t - \beta (V) \left(F_{\rm e}\d x +\, \delta Q\right) \\
-c \, M\,  \beta (V)   \, \d (\gamma v ) + M \, c^2 \, \d \gamma = -\beta (V)\,  c \, (F_{\rm e}-N )\, \d t + F_{\rm e}\d x + \delta Q \end{array} \right. \label{duascoisitas}
\eeq
which is compatible with Equation (\ref{duascois}), as one immediately recognizes.

Had we started with the description in reference frame S$'$,  i.e. with forces not simultaneously applied and net linear momentum associated with the heat,
the transformation $\Lambda_\nu^\mu  (-V)$ would yield, of course, the  description in S, according to the principle of relativity.  On the other hand, if the experiment is conducted in S and correctly described in that reference frame (forces simultaneously applied and thermal reservoir at rest), the transformation $\Lambda_\nu^\mu  (V)$ would automatically provide the  description in S$'$.

\section{Adiabatic inelastic collision}
\label{sec:adiinecoll}

\begin{figure}[hbt]
\begin{center}
\hspace*{0.0cm}
\includegraphics[width=9.0cm]{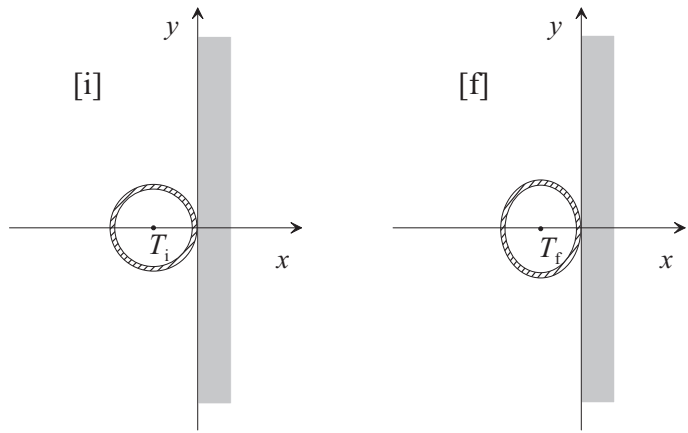}
\end{center}
\vspace*{-0.6cm}
\caption[]{\label{fig:2} \small A ball colliding with a wall at rest in reference frame S in an  adiabatic process.}
\end{figure}

In the previous section we considered the isothermal collision, meaning that there should exist a  heat reservoir with which the system may exchange heat.
In this section we consider an adiabatic collision, i.e. we may imagine a sudden process during which the system does not exchange heat with the surrounding. As such, the system must incorporate the
variation of e.g. kinetic energy that occurs in the process. In relativity this also means that the system must change its inertia. Therefore, $M$ is not a constant parameter always characterizing the system, it rather is a varying function   \cite[p.~264]{sherwood11}. Formally we can imagine that the ball's boundary is an adiabatic one, as represented  in Fig.~\ref{fig:2}.

In the adiabatic process, there is no heat exchange,  $\delta Q^\mu=0$, so the equation (\ref{eqrelat2}) now reduces to
$ \d E^\mu=\delta W^\mu $. On the other hand, in the formalism  developed in section \ref{sec:formalism} and in Appendix A, one has to take proper care of the fact that, for this process, $M$ is not a constant. This means that, on the left hand side of equation
(\ref{explighyt}), which refers to S, one has to perform the following transformations:  $M \ c  \ \d (\gamma v ) \rightarrow  c \ \d (M\, \gamma \, v )$ and $M \ c^2  \ \d (\gamma ) \rightarrow   c^2  \ \d (M \, \gamma )$.
The equation corresponding to (\ref{explighyt}) is now written has
\beq
\left( \begin{array}{c} c \ \d (M\, \gamma \, v ) \\ 0 \\ 0  \\ c^2  \ \d (M \, \gamma ) \end{array} \right) = \left( \begin{array}{c}  c(F_{\rm e}-N) \, \d t\\ 0 \\ 0 \\ F_{\rm e}\, \d x \end{array}\right)\, ,
\label{novadfg}
\eeq
where we have used the fact that the contact force does not perform work. The integration of this equation is straightforward on the left-hand side, because we only have exact differentials (for the three-momentum and for the energy). On the right hand side we may again simplify the approach by considering an average constant braking force, $\bar N$ (the force $F_{\rm e}$ is constant, anyway). We are lead to the following set of equations:
\beq
\left\{ \begin{array}{l}   - c M_{\rm i} \gamma(v_0) v_0 = c\, (F_{\rm e}-\bar N)  t_0 \\ M_{\rm f}\, c^2 - \gamma(v_0) M_{\rm i}\, c^2 = F_{\rm e} \Delta x  \end{array} \right. . \label{duascoisastt}
\eeq
The initial and final temperatures of the body are denoted by $T_{\rm i}$ and $T_{\rm f}$, respectively and the inertia, which is a function of the temperature, $M=M(T)$, is different for the initial and final state:
\beq
M_{\rm f}=M_{\rm i}+ c^{-2} \int_{T_{\rm i}}^{T_{\rm f}} C_P\, \d T \, ,
\eeq
where $C_P$ is the body thermal capacity ($C_P=M\, c_P$, where $c_P$ is the specific heat). As in section 3, we assume that there is no inertia change due to deformations effects of the ball.
From the previous equation and from the second equation (\ref{duascoisastt}) one readily obtains   \cite{redzic06}
\beqa
\Delta M \, c^2&=&  \int_{T_{\rm i}}^{T_{\rm f}} C_P\, \d T \nonumber \\
             &=& [\gamma (v_0) -1] M_{\rm i} c^2 + F_{\rm e} \Delta x \, . \label{dfgeyu}
\eeqa
Under the assumption $M_{\rm i} \approx M_{\rm f}  $  the first equation (\ref{novadfg}) still allows us to write  $M_{\rm i}\,  \d (\gamma \, v )   \approx (F_{\rm e}-N) \, \d t$. By using expression (\ref{er4tsdi})  one concludes that
$M_{\rm i}\, c^2 \,\d (\gamma ) \approx (F_{\rm e}-N) \, \d x$, yielding, after integration,
\beq
M_{\rm i}\, c^2 \, [ 1 - \gamma (v_0) ] \approx (F_{\rm e}-\bar {N}) \Delta x \, .
\eeq
 By comparing this equation with (\ref{dfgeyu}) one arrives at
 \beq
 \Delta M \,   c^2
 \approx \bar N \Delta x
 \eeq
i.e. the inertia increment, which is directly related to the body's internal energy increment,
is also equal to the magnitude of the pseudo-work of the contact force, somehow in analogy with the previous isothermal example where such pseudo-work was equal to the heat flow.

The description of the process in the reference frame S$'$ follows {\em pari passu} the procedure presented in \ref{ssec:prinrelativ}, namely by applying the Lorentz transformation to the matrix equation (\ref{novadfg}). Since the ball's inertia,
at any given instant,  is a Lorentz invariant, both observers agree with the same value for the inertia of the system, in particular, for the initial state, $M_{\rm i}= M(T_{\rm i})$, and
for the final state,  $M_{\rm f}= M(T_{\rm f})$. Therefore they must agree that $T_{\rm i}$ is the same in S and S$'$, and the same happens with $T_{\rm f}$: the temperatures are the same in both reference frames~ \cite{landsberg04}.

\section{Conclusions}
\label{sec:conclus}

The  relation between relativity and thermodynamics is not usually presented in textbooks.
Inspired by the relativistic dynamics for a single particle,  using four-vectors in the Minkowski  space, we generalize that formalism arriving at a suitable one to be  applicable to relativistic  systems of particles, including extended, composite and deformable bodies.

The generalization consists in introducing a
four-vector momentum-energy for an extended  body,  a four-vector for the impulse-work associated with the forces simultaneously applied to the body, and a similar four-vector   associated to the heat, satisfying the maximum entropy principle. These entities obey
an equation that, on the one hand, embodies the conservation of the energy and, on the other, the conservation of the linear momentum.
In the heat four-vector, the fourth component is the energy transfer, and the space-like components are associated to the `non-mechanical impulses' that lead to changes in the three momentum of the system. We keep  a parallelism with the four-vector work, whose fourth component represents the energy exchange with the system as work, the space-like components being the usual impulse of the external resultant force that leads to a variation of the linear momentum of the system.

We applied the formalism to an inelastic collision of a deformable ball, subjected to more than one force, in two different situations: an isothermal process where there is heat exchange with a heat reservoir; and an adiabatic process which results in a change of the temperature and of the inertia of the system itself. The processes are described in frame S in which forces are simultaneously applied and the thermal reservoir (for the isothermal process) is at rest. Then, the Lorentz transformation straightforwardly provides the description of the process in  reference frame S$'$. If the forces in S are simultaneously applied, in the limit $c\, \rightarrow \, \infty$ the non-relativistic descriptions, both in S and in S$'$, are recovered.

\section*{Appendix A}

The metric tensor, $g_{\mu \nu}$, with zero off-diagonal elements, is taken with diagonal elements $(-1,-1,-1,+1)$, where the first three stand for space and the fourth for time.
The norm of a four-vector $A^\mu = ({\vec A}, A_0)$  is $A \equiv \left|\left|A^\mu \right|\right|= (A^\mu A_\mu)^{1/2}= \sqrt{A_0^2-\vec A \cdot \vec A }$,  where $A_\mu = g_{\mu \nu }A^\nu$.
The  position and momentum four vectors, in an obvious shorthand notation, are given by
\beq
x^{\mu}=\left(\begin{array}{c} \vec r  \\ c\, t \end{array}\right)\, , \ \ \
p^{\mu}=\left(\begin{array}{c}  \gamma m \vec v  \\ \gamma m c \end{array}\right)\, .
 \label{eq-34}
\eeq
Usually, the dynamical equation for a single  relativistic particle moving with velocity $\vec v$ in reference frame S, and acted upon by
force $\vec F$, is written in the form given by (\ref{hlop}) that we repeat here for the sake of completeness   \cite{rahaman14}:
\beq
F^\mu = \frac{\d p^\mu}{ \d \tau} \, ,  \label{eq-344}
\eeq
where $\tau$ denotes the proper time measured by the clock that moves with the particle, and
the components of the four-vector  (\ref{eq-344}) are given by
\beq
F^\mu = \left(\begin{array}{c} \gamma \vec F  \\ c^{-1} \, \gamma \, \vec F \cdot \vec v \end{array}\right) \label{eq-544}
\eeq
where $\gamma=\gamma(v)$.  By introducing the four-vector energy, which is the momentum multiplied
by $c$, $E^\mu=c\, p^\mu$, equation  (\ref{eq-344}), after being expressed as $\d E^\mu = c \, \d p^\mu = F^\mu \, c\, \d \tau$, explicitly leads to
\beq
\d E^{\mu}=\left(\begin{array}{c}  c\, m \, \d (\gamma  \vec v)  \\ m \, c^2 \, \d \gamma  \end{array}\right)=
\left(\begin{array}{c}  c \, \gamma \, \vec F \, \d \tau  \\ \gamma  \, \vec F \cdot \vec v \, \d \, \tau  \end{array}\right) \label{energyapp}
\eeq
One should note that $\d t = \gamma \, \d \tau$ is the time interval measured
 by synchronized clocks
in reference frame S. Therefore $\gamma \, \vec v \, \d \tau= \d \vec r$ is the displacement of the force in that frame and the previous equation can be written as
\beq
\d E^{\mu}=
\left(\begin{array}{c}  c \, m  \d ( \gamma  \, \vec  v)  \\ m  c^2 \, \d \gamma  \end{array}\right) =
\left(\begin{array}{c}  c \, \vec F \, \d t  \\   \vec F \cdot \d \vec r   \end{array}\right) = \delta W^\mu \label{xxxyui}
\eeq
where $\delta W^\mu$ is the impulse-work four-vector.
If there are several forces acting upon the particle, we can replace all of them by the resultant force and apply the above derivations to this single resultant force.

However, it is interesting to generalize the fundamental equation  (\ref{xxxyui}) to a finite system (or a system of several particles) with several external forces all acting simultaneously (the internal forces add up to zero).
In this case, one should note that the inertia of the system, $M$, may vary (due to temperature changes, for instance) and there might happen processes involving forces that do not perform work, or involving non-mechanical energy exchanges like heat. The infinitesimal energy variation four vector now should be written as
\beq
\d E^{\mu}=\left(\begin{array}{c}  c  \, \, \d \vec P  \\ \d E \end{array}\right)
\eeq
where the space-like component is the variation of the linear momentum of the system  $\vec P= M\,  \gamma (v) \vec v$   and the time-like component is the variation of the energy, $E= M\gamma (v) c^2  $.
Here we use $\vec v$ to denote the velocity of the system as a whole.
The  energy four-vector for the system is expressed by
\beq
E^{\mu}=c P^\mu = \left(\begin{array}{c}  c\gamma  M(T) \vec v  \\ \gamma  M(T) c^2 \end{array}\right)
 \label{eq-34x}
\eeq
where, by $M(T)$ we already admit that the inertia may vary with the temperature. As explained in subsection 2.1, the inertia of a system is directly related to its internal energy,
$M(T)= c^{-2}\, U(T)$. The equation corresponding to (\ref{internfgt}) for the inertia is (see that equation for the meaning of the quantities in the following one)
\beq
M(T)=\sum_i m_i- \tilde{ U} c^{-2} + M c^{-2} \int _{T_0}^T c_P(T) \, \d T = M+M  c^{-2} \int _{T_0}^T c_P(T)\,  \d T
\eeq
where by $M$ (without any argument) we are denoting the inertia at a reference temperature, $T_0$.
All observers know $M(T)$ since they know the number of elementary particles in the body, their characteristics and how they are organized to form the system. They also know the specific heat $c_P(T)$.
Therefore, at  a given instant, the body's  inertia is defined as
 $M(T)=c^{-2}\left|\left|E^\mu \right|\right| = c^{-2}\left(E^2 - c^2P^2\right)^{1/2}$, which is a relativistic invariant that, of course,  may change along a process.  Because of this invariance, all observers, in different inertial reference frames, assign the same inertia to the system and, consequently, assign also the same temperature.

On the other hand, the impulse-work four vector becomes a summation over the four vector for each external force, namely
\beq
\delta W^\mu= \left(\begin{array}{c}  c \sum_j  \gamma(v_j) \, \vec F_j \ \d \tau_j  \\ \sum_j \gamma(v_j) \, \vec F_j \cdot \vec v_j \ \d \tau_j  \end{array}\right)
\label{ujsdft}
\eeq
where $\d \tau_j$ is the proper time measure by the clock  that travels with  the application point of the $j$-th force that moves with velocity $\vec v_j$ in S (in other words, that clock is located exactly at the application point of the force). Therefore all $\d \tau_j$ are different, in principle, but all $\gamma(v_j) \d \tau_j= \d t$
correspond exactly to the same time interval measured in reference frame S. Hence,  in the reference frame S all forces are  simultaneously applied and this ensures that, in the Newtonian limit $c\, \rightarrow \infty$, they are simultaneously applied too, which is a necessary condition. Notwithstanding, in frame S$'$ these forces will not be simultaneously applied \cite{cavalleri69}.

In conclusion, the  impulse-work four vector in S reduces to
\beq
\delta W^\mu= \left(\begin{array}{c}  c   \vec F_{\rm ext}\d t \\ \sum_j  \vec F_j \cdot \d \vec r _j  \end{array}\right) \, .\label{ystsy}
\eeq
In the space-like component of this matrix equation,  $\vec  F_{\rm ext}= \sum \vec F_j$ is the resultant of the external forces. In the time-like component one has the real work (and not the pseudo-work) of the external forces.

In a certain sense, it can be argued that relativity is most closely related with thermodynamics
  \cite{margaritondo03} than with mechanics   \cite{brown05}.
The equation $\delta E^\mu= \delta W^\mu$ has to be generalized in order to include non-mechanical energy-momentum exchanges with the surrounding. This generalization corresponds, after all, to the implementation in the formalism of the principle of the energy-momentum conservation. It can be written  in the form \cite[p.~303]{sherwood11}
\beq
\d E^\mu = \delta W^\mu + \delta Q^\mu\, ,
\eeq
where $Q^\mu$ is the four vector related to energy exchanges as heat (i.e. that are not mechanical work).
In general, for $Q^\mu$ one has
\beq
Q^{\mu}=\left(\begin{array}{c} c \, \vec P_Q   \\ E_Q \end{array}\right)\label{ghar}
\eeq
and one defines heat as the norm of this four-vector   \cite[p.~94]{rahaman14}, namely
\beq
Q \equiv \left|\left|Q^\mu\right|\right|=\left( E_Q^2 - c^2 \, P_Q^2\right) ^{1/2} \, . \label{calorr}
\eeq
This heat is exchanged between the system and the heat reservoir, which plays an important role.
 For simplicity, let us assume a heat reservoir which is at rest in a given reference frame. The second law of thermodynamics imposes that
the energy exchange with the reservoir should take place with maximum entropy increase of the universe. The entropy variation of the universe is $\Delta S_{\rm U} = {Q\over T}+\Delta S $, where the first term refers to the reservoir and the second term to the body. The body interchanges the energy $E_Q$ which results in a certain fixed value for its entropy variation, $\Delta S$. Hence, the  maximum entropy increase of the universe occurs when, for a given $E_Q$, one has ${\vec P}_Q=0$. This means that, in S, all space components must vanish
in  (\ref{ghar}), as it is the case in the inelastic collision studied in section \ref{sec:ineslascolliso}. In the reference frame S the heat reservoir is at rest and all forces are simultaneously applied. In the same section, we look at the problem from a different inertial reference frame and, indeed,  the space-like components in  (\ref{ghar}) are not zero, $ \vec P_Q \not= 0$. Moreover, in general, the forces are not simultaneously applied, as we explicitly observe in the example treated in subsection 3.1.

We conclude with a remark on the relativistic invariance of $Q$ and $T$ (therefore of the internal energy of the reservoir). As a consequence, the entropy variation of the body and of the universe are relativistic invariants as well. An (infinitesimal) entropy variation can be regarded as the invariant norm of a four-vector defined by $\d S^\mu={\delta Q^\mu \over T}$.

\section*{Appendix B}

The linear momentum and the energy can be expressed as  ${\vec p}= \gamma (v) { M} {\vec v}$ and $E= \gamma (v) M c^2$.
On the other hand, $E^2 = c^2 {\vec p}\cdot {\vec p} + {M}^2 c^4$ and,  by differentiating both sides of this equation, one obtains
$
E \, \d E = c^2 {\vec  p}\cdot  \d {\vec p}\, .
$
Using here the previous expressions for the linear momentum and for the energy, one arrives at an equation equivalent to (\ref{er4tsdi}):
\beq
\d [\gamma (v) c^2] = {\vec v} \cdot \d [\gamma (v) {\vec v}\, ]\, .
\eeq

\section*{Appendix C}

Let us consider a single force
$(F_x, F_y, F_z)$, applied in a certain point of an extended body. This application point moves with velocity
${\vec v}= (v_x, v_y, v_z)$ and its infinitesimal displacement, measured in S, is $\d {\vec r}= {\vec v} \d t = (\d x, \d y, \d z)$, where $\d t$ is the corresponding time interval measured in S (this is not a proper time).
We want to prove that
\beq
\delta W^\mu = \left(\begin{array}{c}  c F_x \d t   \\  c F_y \d t \\ c F_z \d t \\   {\vec F} \cdot \d {\vec r}  \end{array}\right)
\label{fv1}
\eeq
is a four-vector. To this end we must prove that
\beq
\delta {W'}^\mu = \left(\begin{array}{c}  c F'_x \d t'   \\  c F'_y \d t' \\ c F'_z \d t' \\   {\vec F'} \cdot \d {\vec r'}  \end{array}\right)
\label{fv2}
\eeq
is obtained from (\ref{fv1}) by means of a Lorentz transformation
$\delta {W'}^\mu = \Lambda_\nu^\mu \delta W^\nu$, where $\Lambda_\mu^\nu$ is the Lorentz transformation matrix.
We shall only consider the case of the standard configuration of S$'$ relative to S, so that the transformation matrix $\Lambda_\nu^\mu (V)$ is given by
(\ref{eq-1}).

Let us consider (\ref{fv2}) and  the following well-known transformations
\beqa
&& {F'}_x = {F_x - (V/c^2) {\vec F} \cdot  {\vec v} \over 1 - v_xV/c^2}\, ,\, {F'}_y = {{F}_y /\gamma \over 1 - v_xV/c^2} \, ,\, {F'}_z = {{F}_z /\gamma \over 1 - v_xV/c^2} \\
\noalign{\vspace*{0.3cm}}
&&\d x' = \gamma (\d x - V \d t) \, , \, \d y' = \d y \, , \,  \d z' = \d z \, , \, \d t' = \gamma [\d t - (V/c^2) \d x]
\eeqa
where $\gamma=\gamma(V)$. Introducing this primed quantities in  (\ref{fv2}), after some straightforward algebraic manipulations one arrives at
\beqa
c\, {F'}_x \d {t'} &=& c\, \left[\gamma   F_x \d t - {V\over c^2} \gamma {\vec F} \cdot \d {\vec r}\right] \, \\ \nonumber
c\, {F'}_y \d t' &=& c\, F_y \d t \\ \nonumber
c\, {F'}_z \d t' &=& c\, F_z \d t \\
{\vec F'} \cdot \d {\vec r'}  &=& - {V\over c}    \gamma  c F_x \d t  + \gamma  {\vec F} \cdot \d {\vec r}\, . \nonumber
\eeqa
Hence, each component of  $\delta W^\mu$ transforms according to the transformation rule of a four-vector, i.e. $\delta {W'}^\mu = \Lambda_\nu^\mu (V) \delta W^\nu$ for the standard configuration. The same reasoning applies to each force applied to the extended body. Since
the sum of four-vectors is a four-vector, one concludes that  (\ref{ujsdft}) or (\ref{ystsy}) are four-vectors.

It is useful here to borrow the arguments of Gamba in the framework of the asynchronous formulation \cite{gamba67}: one can write a four-vector $A^\mu ({\vec x}, t, {\vec X}, T)$ as the sum of two four-vectors $B^\mu ({\vec x}, t)$ and $C^\mu ({\vec X}, T)$, with the condition $t=T$. In a shorthand notation,
$A^\mu ({\vec x},  {\vec X}, t)= B^\mu ({\vec x}, t) + C^\mu ({\vec X}, t)$.
Because each four-vector $\delta {W}_k^\mu$ ($k=1,2,...,N$) in S refers to exactly the same  time interval, $\d t$, as measured by synchronized clocks in S, the sum
\beq
\delta W^\mu = \sum_k \delta W_k^\mu = \left(\begin{array}{c}  c F_{x, 1} \d t    \\  c F_{y,1} \d t \\ c F_{z, 1} \d t \\   {\vec F}_1 \cdot \d {\vec r}_1  \end{array}\right) +\left(\begin{array}{c}  c F_{x, 2} \d t    \\  c F_{y,2} \d t \\ c F_{z, 2} \d t \\   {\vec F}_2 \cdot \d {\vec r}_2  \end{array}\right)+  \cdots + \left(\begin{array}{c}  c F_{x, N} \d t    \\  c F_{y,N} \d t \\ c F_{z, N} \d t \\   {\vec F}_N \cdot \d {\vec r}_N  \end{array}\right) = \left(\begin{array}{c}  c \sum_k F_{x, k} \d t    \\  c \sum_k F_{y,k} \d t \\ c \sum_k F_{z, k} \d t \\   \sum_k {\vec F}_k \cdot \d {\vec r}_k  \end{array}\right)\, ,
\label{cdf}
\eeq
is a four-vector.  The  sum of these four-vectors defined simultaneously or, better to say, referring to the very same interval of time, is coherent with the classical requirement of simultaneity of all applied forces to the system, acting during the same interval of time, in any reference frame (in the classical limit forces could be applied at different spatial points of the system). By choosing the reference frame S in which forces are simultaneously applied, one ensures the correctness of the classical limit.  Of course,
an observer in S$'$ will observe the same physical situation now with the four-vectors not simultaneously applied.
 Therefore, the impulse-work (\ref{ujsdft}) or (\ref{cdf}) is not a four-vector in the general sense of special relativity because a well-defined four-vector must be defined in each system simultaneously and
it is obtained in another reference frame by a Lorentz transformation. But considering the simultaneity in a particular reference frame, we can consider the impulse-work
as a four-quantity in such reference frame and it can be seen as a four-vector if we applied a Lorentz transformation to such four-quantity. Therefore
four-vector is always related to the reference frame where simultaneity
is defined. When the inverse is done, i.e. when the impulse-work is defined in another reference frame, it will not correspond to the four-quantity obtained by
applying the Lorentz transformation to the four-quantity in the first reference frame.

\end{document}